\documentclass[10pt,conference]{ieeeconf}
\usepackage{blindtext}
\usepackage{graphicx}
\usepackage{CJKutf8}  
\usepackage{caption}
\usepackage{subcaption}
\usepackage{multirow}
\usepackage{amsfonts}
\usepackage{amsmath}
\usepackage{url}
\DeclareMathOperator*{\argmin}{argmin}

\begin{document}

\title{sWSI: A Low-cost and Commercial-quality Whole Slide Imaging System on Android and iOS Smartphones}
\author{Shuoxin Ma*, Tan Wang\\TerryDr Information Technology\\
	Nanjing, Jiangsu, China\\
	Email: sxma@terrydr.com}

\maketitle

\begin{abstract}
In this paper, scalable Whole Slide Imaging~(sWSI), a novel high-throughput, cost-effective and robust whole slide imaging system on both Android and iOS platforms is introduced and analyzed. With sWSI, most mainstream smartphone connected to a optical eyepiece of any manually controlled microscope can be automatically controlled to capture sequences of mega-pixel fields of views that are synthesized into giga-pixel virtual slides. Remote servers carry out the majority of computation asynchronously to support clients running at satisfying frame rates without sacrificing image quality nor robustness. A typical 15x15mm sample can be digitized in 30 seconds with 4X or in 3 minutes with 10X object magnification, costing under \$1. The virtual slide quality is considered comparable to existing high-end scanners thus satisfying for clinical usage by surveyed pathologies. The scan procedure with features such as supporting magnification up to 100x, recoding z-stacks, specimen-type-neutral and giving real-time feedback, is deemed work-flow-friendly and reliable.
\end{abstract}

\begin{keywords}
Mobile health, Image processing, Cloud computing for healthcare, Whole slide imaging
\end{keywords}


\section{Introduction}
Virtual slides generated from whole slide imaging~(WSI) systems is an essential component of digitized diagnostic process, as it provides extended field-of-views(FoVs) under microscopes without handling specimen physically\textsuperscript{\cite{wsi}}. However, the automated scanners that are commonly used to capture and process such data cost approximately \$50,000 or more up-front even for low-frequency usage.

In many developing countries, this financial cost alone has significantly impeded modernizing related departments in hospital, such as the that of pathology in China. Lacking digitization then undermines the productivity and diagnostic accuracy, widely leading to poorer administrative attention and tighter budgets.

In recent years, two alternative solutions have attracted much academic and commercial interest. One is aborting the automation feature thus leaving the operator to control the microscope manually, reducing the product package to a dedicated digital camera and softwares~\textsuperscript{\cite{bg:manual_scan0}\cite{bg:manual_scan1}}, costing as low as \$10,000. The other attempts to make best use of smartphones, which not only have integrated capturing and processing ability but also are widely distributed among clinical professionals thus lowering the start-up cost to near zero. A small number of products in the later category in either research or commercial stage has been evaluated by clinical professionals\textsuperscript{\cite{wsi:iphone}}, but to the limited knowledge of the authors, all of them are made exclusively for high-end iPhones and are not commercially available yet. Although rarely explained explicitly, robustness to guarantee successful virtual slide generation could be a serious obstacle between publishable researches and commercial products. Additionally, diversity in hardware and operating systems might be the reason that Android phones, though dominating handset market in developing countries, are largely ignored. 

In this paper, a WSI system on maintream smartphones just became publicly available with commercial-quality and low cost named scalable WSI~(sWSI)\textsuperscript{\cite{drterry}} is introduced and evaluated. It offers fast and reliable WSI on most handsets, average Androids or flagship iPhones alike, reducing up-front cost to about \$100 and the average service cost per scan is under \$1. Pathologists recognize it as an attractive alternative to stand-alone scanners, especially for quick scans such as with frozen sections as well as medium/low-frequency usages. The rest of this paper is organized as following. In Section~\ref{sec:system}, the overview of system architecture is illustrated. In Section~\ref{sec:failproof}, the client's and server's functions as well as the major techniques to guarantee robustness are analyzed In Section~\ref{sec:correction}, the on-the-fly distortion correction model is formulated with a solution algorithm presented. In Section~\ref{sec:result}, subjective performance evaluations by surveyed clinical users of both automated scanner and sWSI are summarized, which a conclusion drawn in Section~\ref{sec:conclusion}.

\section{System Overview}
\label{sec:system}
There are two essential and costly components in a typical WSI scanners: the capturing unit, typically a set of lens with a distortion-calibrated digital eyepiece, and on-board or external high-performance computers. Like any dedicated devices, since both parts are specifically built for the system, they are mostly non-productive when the system is idle thus waste much of their value when under-used. Unfortunately, this is commonly the case for smaller hospitals where complicated pathological diagnosis occurs but only occasionally. This situation, coupled with consumer electronics' performance approaching medical-grade tools, led to the idea of creating sWSI.
\subsection{Hardware}
\label{sec:system_hardware}
To provide full WSI functionality at a dramatically lower cost, sWSI aims to reduce cost of both hardware necessities. For the optical part, it reversibly upgrades existing microscopes with built-in cameras of smartphones and compensate for the unknown optical distortions computationally, as discussed in detail in Section~\ref{sec:correction}. For the computing part, it utilizes smartphones for light-weight real-time processing and transfers the major bulk to shared remote servers so to allow temporal-multiplexing for improving utilization rate and cost-sharing. 

Even though the prices of mainstream smartphones spreads widely, much of it came in the form of user-friendly features such as security or battery life that is largely irrelevant to sWSI. Thanks to fast expansion of smartphone markets, their cameras, which used to be the critical link in such clinical applications, are now on par with many dedicated digital eyepieces~\textsuperscript{\cite{smartphone_medicine}}. Overall, newer smartphone models can easily meet the minimal requirement listed in Table.~\ref{tab:min_spec} at price as low as \$100. It should be also noted that the higher-end ones that meet the optional specification for better performance may be brought at deep discounts as used or refurbished, which may suffer short battery life or a repaired screen but does not affect the performance of sWSI.
\begin{table}[!h]
	\centering
	\caption{Minimal and Optional Hardware Specifications}
	\begin{tabular}{ | c | c |c |}
		\hline
		Item & Minimal Spec. & Recommended Spec.  \\ \hline 
		OS Version    & Android 4.2 or iOS 9 & N/A \\ \hline
		CPU   & Dual-core @1.2GHz & Quad-core @ 2.4GHz \\ \hline
		Camera & 3MP & 12MP \\ \hline
		Driver & Raw Pixel Data & Exposure control \\ \hline
	\end{tabular}
	\label{tab:min_spec}
\end{table}

\begin{figure}[t]
	\hspace{0cm}
	\centering
	\begin{subfigure}{.25\textwidth}
		\centering
		\includegraphics[trim={0 0 0 0},clip,height = 4cm]{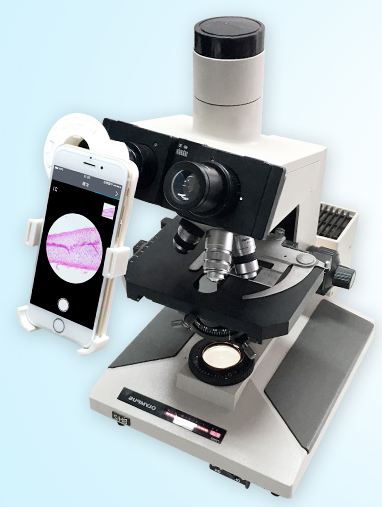}
		\caption{Typical hardware setup}
		\label{fig:hardware}
	\end{subfigure}%
	\hspace{-1cm}
	\centering
	\begin{subfigure}{.25\textwidth}
		\centering
		\includegraphics[trim={0 0 0 0},clip,height = 4cm]{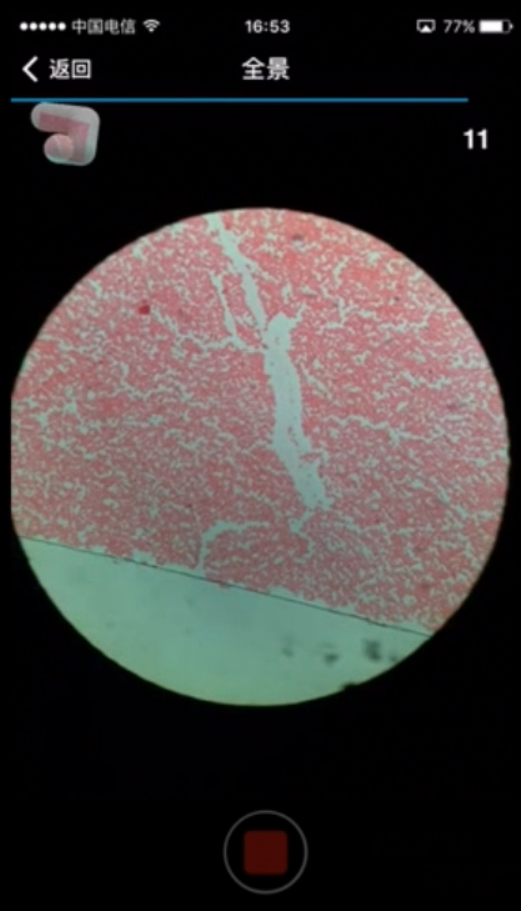}
		\caption{User interface}
		\label{fig:ui}
	\end{subfigure}
	\caption{Hardware and Software User Interface}
	\vspace{-1.5\baselineskip}
	\label{fig:system}
\end{figure}

Considering the fact that most professionals in research and health-care services already own a handset or better ones as specified above, sWSI requires installing only one adapter for each pair of existing smartphone and optical microscope. These microscope-smartphone adaptors are available with many commercial options as well as open-source designs for DIY 3D printing, though the ones specifically built for each phone model are preferred so to minimize need for adjusting camera-eyepiece alignment and to block disruption light sources. One setup is demonstrated in Fig.~\ref{fig:hardware} of an used iPhone 6 costing \$200 installed on Olympus BH2-BHS microscope with scalScope adapter, which took about 15 seconds to set up. \\

\subsection{Software}
\label{sec:system_software}
In addition to image compressing, transferring and virtual slide synthesizing as needed in any whole slide scanning systems, the software in sWSI is also responsible for automatically measuring and compensating hardware diversity.

Unfortunately, fully localizing many of these functions are beyond the reach of mass produced mobile devices. Synthesizing the virtual slide from FoVs requires at least several GB of RAM and sequentially processing hundreds of FoVs at full resolutions can take an hour or more on a mobile CPU. Besides, since the virtual slides will be stored remotely anyway, there is little extra cost in moving the bulk of processing onto remote servers, as implemented in sWSI. The downside of this distributed computing model is introducing significant risks of failure by splitting the processing work-flow into asynchronous ones, but in sWSI this is solved as explained in Section~\ref{sec:failproof}.

Another practical issue worth noticing is that due to architecture and driver support issues beyond the scope of this paper, most Android phones only support JPEG image capture at higher resolution, which cannot be processed pixel-by-pixel. This significantly constrains data flux since each FoV taken has to go through extra encoding-decoding process costing several hundred milliseconds, depending on CPU power and resolution. As a result, the sWSI Android app limits the capturing resolution to about 3MP and generally achieves throughput of about 1 to 3 FoVs per second, except for certain models with drivers offering high-resolution pixel data of images captured, such as the OnePlus X. 

\vspace{-3pt}
\section{Fail-proof Distributed Processing}
\label{sec:failproof}
\subsection{Basic scan procedures and interaction}
\label{sec:failproof_scan}
\begin{figure}[t]
	\centering
	\includegraphics[trim={0 0 0 0},clip,scale = 0.4]{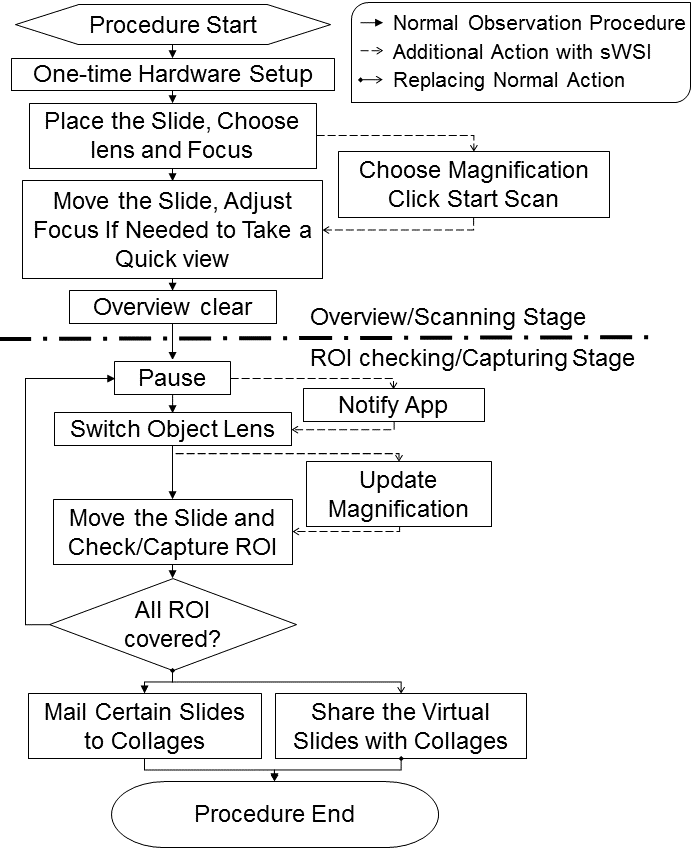}
	\caption{Microscope observation procedures adapted for sWSI}
	\label{fig:procedure}
	\vspace{-0.5\baselineskip}
\end{figure}

In sWSI, a smartphone client app is responsible for gathering user's input, capturing and processing the FoVs as well as guiding users interactively, with a user interface during scan as presented in Fig.\ref{fig:ui}. There is very little difference between the scanning procedure with sWSI as shown in Fig.~\ref{fig:procedure}, and that practised by most microscope users except for requiring updating magnifications, thus is not further discussed here. 

\subsection{Real-time feedback on Clients}
\label{sec:failproof_client}
The client's share of processing focuses on speed and robustness instead of accuracy thus uses down-sampled copies of camera input. It roughly estimate pairwise translation of FoVs by stitching each captured FoV with the last one through key point detection and matching with the SURF algorithm\textsuperscript{\cite{surf_app}}. This translation is then used in three ways: updating a mini-map illustrating current location on the slide, feeding a finite-state machine to manage the kernel asynchronously and providing feedback to users as guidance for operating the microscope. The feedback and their triggers include:
\subsubsection{Moving too fast}
The translation is too far so the key point matching in SURF may be unreliable.
\subsubsection{Lost}
No reliable translation can be obtained. The causes cannot be further distinguished by the machine but should be noticeable to the users, such as moving so fast that there is little overlapping between the current pair of FOVs or the camera is out of focus.
\subsubsection{Touching a boundary}
There are few key points detected so the FoV is likely near a boundary.
\subsubsection{No error}
The translation is reliable. 

With users following the hints, sWSI essentially creates a closed feedback loop that allow scan-time interference against potential failure, such as inability to focus properly on thick samples or to track positioning on barren ones. This mechanism thus prevents most flops due to sample preparation and user operation before spending long time in completing the scan, which is a common issue with automatic scanners.

\vspace{-3pt}
\subsection{Full resolution processing with a-priori knowledge on Servers}
\label{sec:failproof_server}
The cloud servers in sWSI are the primary powerhouses of computation. With full resolution FoVs and scan results from clients, servers re-stitch the adjacent FoVs at maximal accuracy, correct distortion and generate the virtual whole slide. The asynchronous two-staged stitching performed respectively on the clients and servers, however, has inherent weak spots on both stability and efficiency. 

On one hand, the FoV pairwise stitching is based on key point detection and matching, whose outcome in turn is resolution-dependent. As a result, such outcomes in down-sampled and original resolution may potentially be significantly inconsistent. In many cases, as in almost every virtual slides constructed from 100 FoVs or more with prototypes of sWSI, the full-resolution stitching produce unreliable matching at least once or more. 

On the other hand, by the law of large numbers, it is desirable to match as many key point pairs as possible for accurate estimation of the FoV-wise matching function, especially where this function has high degrees of freedom as is the case of sWSI where raw images are non-linearly distorted in unknown patterns. The computational cost of brute-and-force key point matching, however, grows quadratically with number of key points. 

To encounter both issues at once, sWSI employs a a-priori-knowledge-based SURF KP detection and matching algorithm on the server. Recall that SURF detects KPs from a virtual image pyramid that has lower resolution on higher layers. In sWSI, instead of detecting with one threshold across all layers, multiple thresholds are chosen adaptively as following. First, one threshold $t_{r_{ds}}$ is picked to ensure at least $p_hn_{kp}$ KPs are detected on layers $[l_{ds},l_{max}]$ in total, where $p_h\geq1$ is a constant multiplier, $n_{kp}$ is the number of KP detected in the down-sampled copy during scan, $l_{max}$ is the index of the upper most layer and $l_{ds}$ is derived from the power-of-two down-sample rate during scan $r_{ds}$ as
\begin{equation}
	 l_{ds}=\log_2{r_{ds}}, l_{ds}\in{N}.
\end{equation}
Next, threshold $t_i, i\in[0,l_{ds})$ for detection on layer $i$ is adaptively chosen so that $p_ln_{kp}$ KPs are detected on each lower layers, where $p_l>0$ is another constant. With this thresholding approach, most KPs on the scan stage can be detected at full resolution with additional ones from lower layers that are more localized but with higher resolution, while the total number of KPs is controlled by $p_h$ and $p_l$ thus would not over-expand. 

Afterwards, instead of brute-and-force matching by calculating difference of all pairs of KPs and picking the optimal set of match, sWSI selectively calculate those within a constant distance from the coordinates indicated by the scan stage translation with up-sampling and assume all others infinitely large. Assuming there are $m_{kp0}$ and $m_{kp1}$ KPs detected respectively on the pair of FoVs and approximately $k$ other KPs within each KP's selected matching region, a brute-and-force matching needs to calculate differences of $m_{kp0}m_{kp1}$ KP pairs while the proposed methods does so only $(m_{kp0}+m_{kp1})k$ times. Considering that the $m_{kpi}$ is in the range of thousands while $k$ is usually under 10, the proposed modification dramatically reduces calculation yet yields nearly identical results. Since KP matching takes a large number of float point operations thus consumes a large portion of time, this reduction boosts the overall efficiency of sWSI by over $50\%$.

\section{On-the-fly Image Distortion Correction}
\label{sec:correction}
When stitching each FoV pair to match KPs, the projection function can be in any format so long as it minimize error without over-fitting. Combining all FoVs into a single continuous view, however, requires the projection to be linear thus the non-linear distortion must be corrected first. If not, the order of the stacked-up non-linear transfer function of each FoV onto the whole slide will keep growing by each FoV and become very inefficient to solve.

Designed to fit any combination of microscope and smartphone models, sWSI assumes a generalized high-order polynomial~(HOP) inverse-distortion model~\textsuperscript{\cite{distortion}}, which mathematically approximates any function with marginal error if the order is sufficiently high, as proven by the Taylor's theorem. Specifically, it is assumed that there exists a constant but unknown HOP projection function for each scan procedure that maps the raw FoVs into a corrected 2D space, where any matched pixel pairs in overlapping FoVs share the same phase difference for that FoV pair. In another word, after the raw FoVs are corrected by a HOP function, each adjacent FoV pair can be stitched with just translation onto each other with small error. In sWSI, this HOP projection matrix is solved iteratively based on FoV-pair-wise KP matching, formulated as following. 

First, assume the HOP model has $O_p$ orders. Also name the two FoVs in the $i$th FoV pair as $\text{srcFoV}_i$ and $\text{dstFoV}_i$, whereas $\text{srcFoV}_i$ is stitched onto $\text{dstFoV}_i$. For point $j$ on $\text{srcFoV}_i$ with a 2D coordinate $\hat{x}_{i,j}=[x_{i,j,1}, x_{i,j,2}]$, its polynomial expansion kernel is derived as~\textsuperscript{\cite{distortion}}
 $\bar{x}_{i,j} = \bar{\phi}(x_{i,j,1}, x_{i,j,2})$
, where 
\begin{equation}
\bar{\phi}_{O_p}(u,v) = (1,u,v,u^2,uv,v^2,\dots,v^{O_p})
\end{equation}
thus with the number of dimensions being $N_e=\frac{(O_p+1)(O_p+2)}{2}$. Similarly, the point's exact match in $\text{dstFoV}_i$ has a coordinate and kernel $\hat{y}_{i,j}$ and $\bar{y}_{i,j}$, respectively. For simpler notations, also define $N_{e-1}=N_e-1$.\\
Next, note the correction projection matrix as $\boldsymbol{\beta}\in{\mathbb{R}^{N_e \times 2}}$. The linear projection used to stitch the pair after correction is an affine one in the form of $\boldsymbol{A}_i = [\bar{T}_i^T,\boldsymbol{R}_i^T]^T$, where $\bar{T}_i\in{\mathbb{R}^{1 \times 2}}$ and $\boldsymbol{R}_i\in{\mathbb{R}^{2 \times 2}}$ are the translation and rotation components, respectively. The whole model would ideally satisfy
\begin{equation}
\bar{x}_{i,j}\boldsymbol{\beta}\boldsymbol{R}_i+\bar{T}_i-\bar{y}_{i,j}\boldsymbol{\beta} = 0
\end{equation}
for all point pairs across all FoV pairs, where $\boldsymbol{\beta}$ is constant and $[\bar{T}_i,\boldsymbol{R}_i]$ are FoV-pair-dependent but point-pair-independent.\\
In reality, correction error exists and the process turns into solving a constrained optimization problem 
\begin{equation}
 \label{eq:op_prbm}
 \boldsymbol{\beta},{{\{\bar{T}_i,\boldsymbol{R}_i\}}}
 =
 \argmin_{\boldsymbol{\beta},{{\{\bar{T}_i,\boldsymbol{R}_i\}}}} \sum_{i}\mathbb{S}_i(\boldsymbol{\beta},\bar{T}_i,\boldsymbol{R}_i),
\end{equation}
where
\begin{equation}
\begin{split}
 \mathbb{S}_i(\bar{\beta},\bar{T}_i,\boldsymbol{R}_i)
 =
 \sum_{j}
 \Big[||\bar{x}_{i,j}\boldsymbol{\beta}\boldsymbol{R}_i+\bar{T}_i-\bar{y}_{i,j}\boldsymbol{\beta}||^2+\\
 \lambda||\bar{x}\boldsymbol{\beta}-\hat{x}_{i,j}||^2\Big].
\end{split}
\end{equation}
The term $\lambda||\bar{x}\boldsymbol{\beta}-\hat{x}_{i,j}||^2$ here prevents the projections from collapsing into all zeros and $\lambda = 0.001$ is used. It should be noted that based on the assumption that $\boldsymbol{\beta}$ should correct and only correct non-linear distortions, the elements in
\begin{equation}
	\begin{matrix}
	\boldsymbol{\beta} 
	&=&
	\begin{bmatrix}
	\beta_{00},\beta_{10},\dots,\beta_{N_{e-1}0}\\
	\beta_{01},\beta_{11},\dots,\beta_{N_{e-1}1}
	\end{bmatrix}^T
	\\
	&=&
	\begin{bmatrix}
	\beta_{00},\beta_{10},\beta_{20},\tilde{\beta}_0\\
	\beta_{01},\beta_{11},\beta_{21},\tilde{\beta}_1
	\end{bmatrix}^T
	\end{matrix}
\end{equation}
satisfy 
\begin{equation}
 \beta_{00}=\beta_{01}=\beta_{20}=\beta_{11}=0, \beta_{10}=\beta_{21}=1
\end{equation}
Then, the multi-variable non-linear equation of Eq.~\ref{eq:op_prbm} can be solved by iteratively fixing either $\boldsymbol{\beta}$ or ${\bar{T}_i,\boldsymbol{R}_i}$ and find the least-mean-square solution of the other until convergence. Specifically, by freezing $\boldsymbol{\beta}$, 
\begin{equation}
[\bar{T}_i^T,\boldsymbol{R}_i^T]^T = 
(\sum_{j}(\bar{y}_{i,j}\boldsymbol{\beta}\check{x}_{i,j}^T))
(\sum_{j}(\check{x}_{i,j}\check{x}_{i,j}^T))^{-1}
\end{equation}
where
$\check{x}_{i,j} = [1,\bar{x}_{i,j}\boldsymbol{\beta}$]. By keeping $\{\bar{T}_i,\boldsymbol{R}_i\}$ constant and splitting the elements as
\begin{equation}
\bar{T}_i=
\begin{bmatrix} 
T_{i,0}& T_{i,1} 
\end{bmatrix}
,
\boldsymbol{R}_i=
\begin{bmatrix} 
R_{i,00} & R_{i,01} 
\\
R_{i,01} & R_{i,11}
\end{bmatrix}
\end{equation}
variable elements in $\boldsymbol{\beta}$ can be solved as
\begin{equation}
\begin{bmatrix} 
\tilde{\beta}_0
\\
\tilde{\beta}_1
\end{bmatrix}
=
\sum_{i,j}((\boldsymbol{M}_{i,j}^T\boldsymbol{M}_{i,j})^{-1})
\sum_{i,j}(\boldsymbol{M}_{i,j}^T\tilde{L}_{i,j})
\end{equation}
where
\begin{equation}
\label{eq_mid}
\begin{matrix}
&\boldsymbol{M}_{i,j}&
=&
\begin{bmatrix} 
\boldsymbol{M}_{i,j,00}(\tilde{x}, \tilde{y}) & \boldsymbol{M}_{i,j,01}(\tilde{x}, \tilde{y})
\\
\boldsymbol{M}_{i,j,10}(\tilde{x}, \tilde{y}) & \boldsymbol{M}_{i,j,11}(\tilde{x}, \tilde{y})
\end{bmatrix}
\\
&\tilde{L}_{i,j}&
=&
\begin{bmatrix} 
\tilde{L}_{i,j,0}(\tilde{x}_{i,j}, \tilde{y}_{i,j})\\
\tilde{L}_{i,j,1}(\tilde{x}_{i,j}, \tilde{y}_{i,j})
\end{bmatrix}.
\end{matrix}
\end{equation}
Omitting subscription $(\bullet)_{i,j}$ and $(\bullet)(\tilde{x}_{i,j}, \tilde{y}_{i,j})$ for simplicity, elements in Eq.\ref{eq_mid} are calculated from
\begin{equation}
\begin{matrix} 
\boldsymbol{M}_{00}=
(R_{00}\tilde{x}-\tilde{y})^T(R_{00}\tilde{x}-\tilde{y})+
R_{01}^2\tilde{x}^T\tilde{x}+
\lambda\tilde{x}^T\tilde{x}
\\
\boldsymbol{M}_{01}=
R_{10}(R_{00}\tilde{x}-\tilde{y})^T\tilde{x}+
R_{01}\tilde{x}^T(R_{11}\tilde{x}-\tilde{y})
\\ 
\boldsymbol{M}_{10}=
R_{10}\tilde{x}^T(R_{00}\tilde{x}-\tilde{y})+	
R_{01}(R_{11}\tilde{x}-\tilde{y})^T\tilde{x}
\\
\boldsymbol{M}_{11}=
R_{10}^2\tilde{x}^T\tilde{x}+	
(R_{11}\tilde{x}-\tilde{y})^T(R_{11}\tilde{x}-\tilde{y})+
\lambda\tilde{x}^T\tilde{x}
\end{matrix}
\end{equation}
\begin{equation}
\begin{matrix}
\tilde{L}_0=-\tilde{T}_0(R_{00}\tilde{x}-\tilde{y})^T-R_{01}\tilde{T}_1\tilde{x}^T
\\
\tilde{L}_1=-R_{10}\tilde{T}_0\tilde{x}^T-\tilde{T}_1(R_{11}\tilde{x}-\tilde{y})^T
\end{matrix}
\end{equation}
where $\tilde{x}=\tilde{\phi}_{O_p}(x_{1},x_{2})$ and $\tilde{y}=\tilde{\phi}_{O_p}(y_{1},y_{2})$ are respective sub-vectors of $\bar{x}$ and $\bar{{y}}$ calculated as 
\begin{equation}
\tilde{\phi}_{O_p}(u,v) = (u^2,uv,v^2,u^3,u^2v,\dots,v^{O_p}).
\end{equation}
From experiments, it is shown that $O_p\ge7$ is generally sufficient and the model takes about 10 iterations to converge.
\section{Results and Evaluations}
\label{sec:result}
Currently, sWSI services are in open beta test for iPhones internationally with a special version in P.R.China for both Android and iPhones. Due to national Internet gateway issues, using the version on the opposite side of the boarder may experience slower connection. Sample slides produced by both Android and iPhones can be accessed on the products homepage\textsuperscript{\cite{drterry}}, with data available for further evaluation by contacting the authors.

\begin{figure}[]
	\centering
	\includegraphics[trim={0 0 0 0},clip,scale = 0.3]{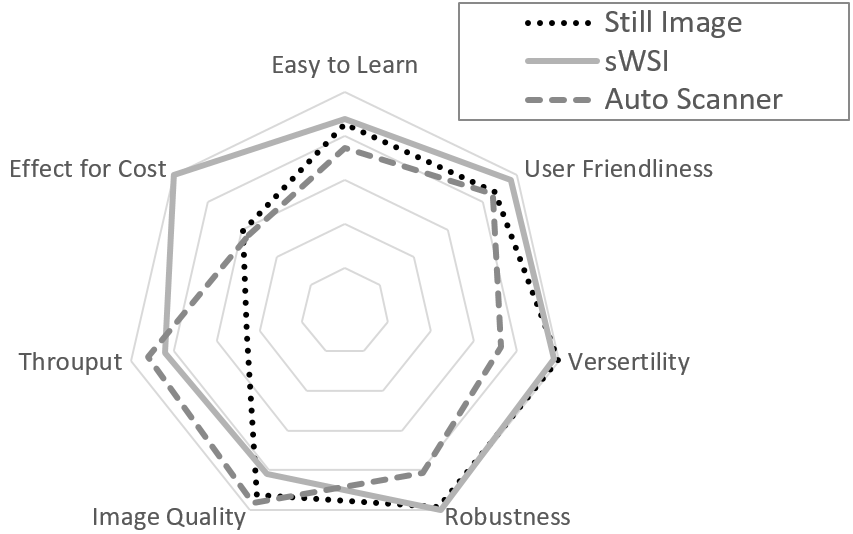}
	\caption{Average Score in Survey}
	\label{tab:survey}
	\vspace{-1\baselineskip}
\end{figure}

The sWSI systems have been used by both trained pathologists and assistant technicians to scan hundreds of samples with satisfaction, sometimes preferred over automatic scanners for versatility and robustness. 30 of sWSI users~(including 22 pathologists) from 8 hospitals in Shanghai, Nanjing, and Luoyang, P.R.China are randomly chosen and continuously followed for at least one month to complete a survey. Most of them report about throughput of 1 FoV per second after just using it for a few times and no failure encountered as long as they operated properly. Statistics of their rating on sWSI experience compared to taking still images and using automatic scanners is summarized in Table~\ref{tab:survey}, with a score scale of 1~(Poor) to 5~(Perfect). It should be noted that many of these respondents uses automatic scanners almost daily, thus may have overrated the cost-for-effect of them as compared to those work in low-level hospitals who do so much less often.
Yet, there are a few issues to be solved in future research and development. Firstly, the FoVs are often non-evenly illuminated and leaves noticeable brightness discontinuity in the virtual slide, possibly caused by improper installation of handsets on the adapter or poor light source. Secondly, some parameters on Android phones cannot be controlled through API for older Android OS. Weakened control may lead to improper configuration, such as a long exposure time causing blur. Lastly but not least, the openGL driver which offers GPGPU computing potential, are very tricky to work with and produces unexpected results on many smartphone models for reasons unknown. Preliminary research using GPGPU on iPhones brought a dramatic boost in processing speed over 60\%, but older models upgraded to iOS10 no longer works properly.

\section{Conclusion}
\label{sec:conclusion}
In this paper, an ultra-low-cost whole slide imaging system with client hosted on mainstream Android and iOS smartphones is introduced and analyzed. Compared to automatic scanners and high-end-computer-based solutions, this alternative dramatically reduces the setup cost to as low as \$100 per unit with service cost under \$1 per scan.

By employing distributed image processing, both robustness and efficiency are covered. Through high performance computing and realtime feedback, user friendliness is optimized with minimal manual input, leaving most interface-kernel coordination and even image distortion correction fully automated. 30 surveyed clinical professionals give sWSI a higher score on most aspects as compared to automatic scanners, except for a slightly poorer image quality and lower throughput.

\end{document}